# Complexity of Inference in Graphical Models

**Venkat Chandrasekaran**
Laboratory for Information and Decision Systems
Massachusetts Institute of Technology
Cambridge, MA 02139

**Nathan Srebro, Prahladh Harsha**
Toyota Technological Institute – Chicago
Chicago, IL 60637

## Abstract

It is well-known that inference in graphical models is hard in the worst case, but tractable for models with bounded treewidth. We ask whether treewidth is the only structural criterion of the underlying graph that enables tractable inference. In other words, is there some class of structures with unbounded treewidth in which inference is tractable? Subject to a combinatorial hypothesis due to Robertson et al. (1994), we show that low treewidth is indeed the *only* structural restriction that can ensure tractability. Thus, even for the "best case" graph structure, there is no inference algorithm with complexity polynomial in the treewidth.

## 1 INTRODUCTION

Graphical models offer a convenient representation for joint probability distributions over a large number of variables. Such models are defined as stochastic processes with respect to a graph: each vertex of the graph is associated with a random variable, and the edge structure specifies the conditional independence (Markov) properties among the variables. Due to their powerful modeling capabilities, graphical models have found applications in a variety of fields including computer vision (Szeliski, 1990), coding theory (McEliece et al., 1998), and statistical physics (Parisi, 1988). In many of these applications a commonly encountered problem is that of estimation or *inference*, which refers to computing the posterior marginal distribution of a variable at some vertex. We study the computational complexity of the inference problem in a graphical model consisting of discrete-valued variables as a function of structural properties of the underlying graph such as treewidth and planarity.

It is well-known that inference is **NP**-hard if no assumptions are made about the structure of the underlying graphical model (Cooper, 1990), and remains **NP**-hard even to approximate (Roth, 1996) — assuming $\mathbf{P} \neq \mathbf{NP}$, for any algorithm there are some structures in which (approximate) inference takes time super-polynomial in the size of the structure. However, inference in specific structures can still be tractable. For models in which the underlying graph has low treewidth, the junction-tree method provides an effective inference procedure that has complexity polynomial in the size of the graph, though exponential in the treewidth.

The notion of treewidth (Robertson and Seymour 1983; 1986) has led to several results in graph theory (Robertson et al., 1994) and to practical algorithms for a large class of **NP**-hard problems (Freuder, 1990). Among these problems is inference in graphical models, which, as mentioned earlier, can be solved in polynomial-time if the treewidth of the underlying graphs is bounded. For a triangulated graph the treewidth is one less than the size of the largest clique, while the treewidth of a general graph is the smallest treewidth over all triangulations of the graph.

A variety of results focus on special conditions on the potential functions (Fisher, 1966) that result in inference being tractable. Recently, some authors have attempted to provide *structural* conditions, aside from low treewidth, that allow efficient inference. For example, Chertkov and Chernyak (2006) describe an inference method based on enumeration over "generalized loops", suggesting that inference is tractable in graphs with a relatively small number of such loops.

In this paper we consider whether there might be an alternate structural property of graphs, which does not imply low treewidth, but guarantees tractable inference. Put differently we investigate the "best case", rather than worst case, hardness of inference with respect to the treewidth: does inference remain hard even in the "easiest" high-treewidth graph structures?

We focus only on structural properties, and consider algorithms that work (and are tractable) for any choice of the potentials.

Recently, Marx (2007) showed that constraint satisfaction problems (CSP) defined on any class of graphs with unbounded treewidth cannot be solved in polynomial-time. However, Marx's result refers only to algorithms for CSPs involving variables of *unbounded cardinality* (i.e. an unbounded number of states). Allowing such high-cardinality variables plays a critical role in the proof, which employs constructions using models in which the cardinality of the variables grows in an unbounded manner with the size of the graphs. Thus, translating these results to problems of inference in graphical models is of limited interest for typical inference problems. Marx's result can only imply hardness of inference when the number of states for each variable is unbounded, while most inference problems of interest involve variables with low cardinality or even binary states. Indeed, we usually think of the complexity of inference, or even of representation of a discrete graphical model, as growing exponentially with the number of states.

We focus on the complexity of inference in models consisting of *binary* variables defined on any class of graphs with unbounded treewidth. In such models a hardness result can be obtained if we assume a well-known hypothesis from graph minor theory. A minor of a graph $\mathcal{G}$ is a graph $\mathcal{H}$ that can be obtained from $\mathcal{G}$ by a sequence of vertex/edge deletions and/or edge contractions (see Section 2.4 for a precise definition). In a series of over twenty papers, Robertson and Seymour shed light on various aspects of graph minors and proved important results in graph theory. The theorem of greatest relevance to this paper is one that relates graph minors and treewidth: for each $g \times g$ grid-structured graph $\mathcal{G}$, there exists a finite $\kappa_{\text{GM}}(g)$ such that $\mathcal{G}$ is a minor of *all* graphs with treewidth greater than $\kappa_{\text{GM}}(g)$. The best known lower-bound and upper-bound for $\kappa_{\text{GM}}(g)$ are $\Omega(g^2 \log g)$ and $2^{O(g^5)}$ respectively. The *grid-minor hypothesis* holds that $\kappa_{\text{GM}}(g)$ is polynomial bounded with respect to $g$. The hypothesis is based on the belief that $\kappa_{\text{GM}}(g)$ is closer to $\Omega(g^2 \log g)$ than $2^{O(g^5)}$ (Robertson et al., 1994); further evidence in support of this hypothesis is provided in Demaine et al. (2006).

**Main result** There is no class of graphical models consisting of *binary* variables with unbounded treewidth in which inference can be performed in time polynomial in the treewidth. The assumptions behind this result are that $\mathbf{NP} \not\subseteq \mathbf{P/poly}$[1] and the grid-minor hypothesis. Consequently, for *every* sequence of graphs $\{\mathcal{G}_k\}_{k=1}^{\infty}$ indexed by treewidth, inference is super-polynomial with respect to the treewidth $k$. More precisely, for *every* sequence of graphs $\{\mathcal{G}_k\}_{k=1}^{\infty}$ indexed by treewidth, there exists a choice of potential functions such that inference is super-polynomial with respect to treewidth $k$.

We also show the above result for planar graphs without requiring recourse to the grid-minor hypothesis — assuming *only* that $\mathbf{NP} \not\subseteq \mathbf{P/poly}$, there is no class of graphical models with binary variables defined on planar graphs with unbounded treewidth in which inference has complexity polynomial in the treewidth. We obtain sharper versions of these results that are based on the so-called exponential-time hypothesis (Impagliazzo et al., 2001) rather than the assumption that $\mathbf{NP} \not\subseteq \mathbf{P/poly}$. We further extend the hardness result above to hardness of approximation of the partition function to within an additive constant by a randomized polynomial-time algorithm.

**Proof overview** The standard "worst-case" hardness for the inference problem shows that there is *some* family of graphs $\{\mathcal{H}_k\}_{k=1}^{\infty}$ for which the inference problem is hard. In fact, it is known that the family of graphs can be assumed to be planar. To prove "best-case" hardness, we need to show that inference is hard in *every* family of graphs $\{\mathcal{G}_k\}_{k=1}^{\infty}$ of increasing treewidth. We prove this by showing that given *any* family of graphs $\{\mathcal{G}_k\}$ of increasing treewidth, we can reduce the general inference problem on planar graphs to the inference problem on $\{\mathcal{G}_k\}$. Our main tools are the Robertson-Seymour graph-minor theorems, which show that any planar graph is a minor of a not too large grid and a grid is a minor of any graph with not too large treewidth. We use these results to reduce the inference problem on any planar graph to one on a grid of not too large size and then to any graph (in particular $\mathcal{G}_k$) of not too large treewidth ($k$). Unfortunately, the known theoretical guarantees on the "not too large treewidth" is too weak (in fact, super-exponential) for our purposes. We get around this problem by either relying on the grid-minor hypothesis which conjectures that the "not too large treewidth" is in fact at most polynomial in the size of the grid, or by assuming that $\{\mathcal{G}_k\}$ is a family of planar graphs. This completes our proof but for one caveat: it is not sufficient if we know that any planar graph is a minor of the grid and the grid is a minor of $\mathcal{G}_k$; we actually need the sequence of minor operations between these graphs. In general, finding the sequence of minor operations that transform one arbitrary graph to another is an $\mathbf{NP}$-complete problem. However, it is known that a planar graph can be embedded in a grid in linear time (Tamassia & Tollis, 1989). This solves one of our two problems. For the other (embedding a grid into $\mathcal{G}_k$), we exploit the fact that both these graphs are fixed and depend only on the size of the input instance and not on the actual instance itself. Hence, we could non-uniformly hardwire this sequence of minor operations into our algorithm and thus obtain a non-uniform hardness reduction.

---

[1] The assumption $\mathbf{NP} \not\subseteq \mathbf{P/poly}$ is the non-uniform version of the more popular $\mathbf{NP} \neq \mathbf{P}$ assumption. For more details on uniform vs. non-uniform algorithms, see Section 2.2.

**Organization** The rest of this paper is organized as follows. Section 2 provides a brief background on inference in graphical models, treewidth, and graph minors. Section 3 presents the formal statement of the problem addressed in this paper. Section 4 describes constraint satisfaction problems; we prove a reduction from such problems to inference in graphical models that plays a key role in our analysis. Section 5 provides the main results of this paper. We conclude with a brief discussion and open questions in Section 6.

## 2 BACKGROUND

### 2.1 GRAPHICAL MODELS AND INFERENCE

A *graph* $\mathcal{G} = (V, \mathcal{E})$ consists of a set of vertices $V$ and associated edges $\mathcal{E} \subset \binom{V}{2}$, where $\binom{V}{2}$ is the set of all unordered pairs of vertices. A *graphical model* (Lauritzen, 1996) is a collection of random variables indexed by the vertices of a graph; each vertex $v \in V$ corresponds to a random variable $x_v$, and where for any $A \subset V$, $x_A = \{x_v | v \in A\}$. We assume that each of the variables $x_v$ is discrete-valued with cardinality $q$. Of interest in this paper are distributions that factor according to a graph $\mathcal{G} = (V, \mathcal{E})$ as follows:

$$p(x_V) = \frac{1}{Z(\psi)} \prod_{v \in V} \psi_v(x_v) \prod_{E \in \mathcal{E}} \psi_E(x_E). \quad (1)$$

Here, each $\psi_E$ (or $\psi_v$) is only a function of the variables $x_E$ (or variable $x_v$). The functions $\psi_v$ and $\psi_E$ are non-negative and are also known as *potential* or *compatibility* functions. We denote the collection of these potentials by $\psi = \{\psi_v, v \in V\} \cup \{\psi_E, E \in \mathcal{E}\}$. The function $Z(\psi)$ is called the *partition* function and serves to normalize the distribution:

$$Z(\psi) = \sum_{x_V \in \{0,\cdots,q-1\}^{|V|}} \prod_{v \in V} \psi_v(x_v) \prod_{E \in \mathcal{E}} \psi_E(x_E). \quad (2)$$

Given a posterior distribution that factors according to a graph as described above, a common task in applications such as image processing and computer vision (Szeliski, 1990) is to compute the marginal distribution of a variable at some vertex. It is well-known that the complexity of computing the marginal distribution at some vertex is comparable to that of computing the partition function. A polynomial-time procedure to solve one of these problems can be used to construct a polynomial-time algorithm for the other. Thus, we consider in this paper the complexity of computing the partition $Z(\psi)$, and with an abuse of terminology, it is this problem that we refer to as *inference*. The intractability of inference arises due to the fact that there are exponentially many terms in the sum in (2). We study the complexity of inference as a function of structural properties of the underlying graph.

### 2.2 UNIFORM VS. NON-UNIFORM ALGORITHMS

The classical notion of algorithms refers to "uniform algorithms" in which one has a single algorithm that works for all input lengths. A "non-uniform algorithm" on the other hand refers to a family of algorithms, one for each input length. Another way of thinking about such non-uniform algorithms is that the algorithm is allowed to receive some arbitrary oracle advice that depends only on the input length (but not on the actual input). In the theory of computation literature, such "non-uniform algorithms" are usually referred to as fixed-input-size "circuits", where for each input length a different circuit is used. The class **P** is the class of problems that have polynomial time uniform algorithms while **P/poly** is its non-uniform counterpart, i.e., the class of problems that have polynomial time non-uniform algorithms (circuits). Clearly, $\mathbf{P} \subset \mathbf{P/poly}$. The non-uniform version of the assumption $\mathbf{NP} \neq \mathbf{P}$ is $\mathbf{NP} \not\subseteq \mathbf{P/poly}$, and (though slightly weaker) is equally believed to be true. We need to work with the latter assumption since our proof proceeds by reducing the "best-case" inference problem to a non-uniform algorithm for **NP**.

### 2.3 GRAPH TREEWIDTH

A graph is said to be *triangulated* if every cycle of length greater than three contains an edge between two non-adjacent vertices. The *treewidth* tw($\mathcal{G}$) of a triangulated graph $\mathcal{G}$ is one less than the size of the largest clique. The treewidth of a general graph is defined

$$\text{tw}(\mathcal{G}) = \min_{\mathcal{H} \supseteq \mathcal{G}, \mathcal{H} \text{ triangulated}} \text{tw}(\mathcal{H}).$$

Here, $\mathcal{H} \supseteq \mathcal{G}$ denotes that $\mathcal{H}$ is a supergraph of $\mathcal{G}$. In words, the treewidth of a graph $\mathcal{G}$ is the minimum over the treewidths of all triangulated supergraphs of $\mathcal{G}$.

Figure 1 shows an example of a non-triangulated graph $\mathcal{G}$, which has a 4-cycle with no edge connecting non-adjacent vertices. The graph $\mathcal{H}$ is a triangulated supergraph of $\mathcal{G}$, and has treewidth 3 as the largest clique is $\{2, 5, 6, 8\}$. Thus, the treewidth of $\mathcal{G}$ is also 3.

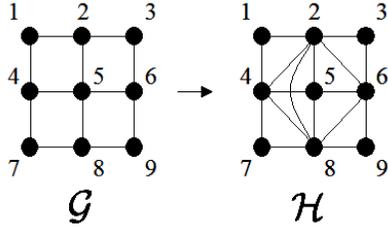

Figure 1: A non-triangulated graph $\mathcal{G}$, and a triangulated supergraph $\mathcal{H}$ of $\mathcal{G}$.

The complexity of a graphical model is often measured by the treewidth of the underlying graph. Distributions defined on trees, which are treewidth-1, permit very efficient linear-time inference algorithms. For loopy graphs that have low treewidth the junction-tree method (Cowell et al., 1999) provides an efficient inference algorithm. However, for general loopy graphs the junction-tree method might be intractable because it scales exponentially with the treewidth. As a result considerable effort is being devoted to the development of approximate inference algorithms. Our focus here is on the computational complexity of exact (or near-exact) inference.

### 2.4 GRAPH MINORS

The theory of graph minors plays a key role in our analysis. Specifically, we show in Section 2.4.1 that the complexity of inference in a minor of $\mathcal{G}$ is bounded by the complexity of inference in $\mathcal{G}$. A *minor* of a graph is obtained by any sequence of the following operations:

- **Vertex deletion**: Given a graph $(V, \mathcal{E})$, a vertex $v \in V$ is *deleted*, as are all the edges $\mathcal{E}_v = \{E \in \mathcal{E} : v \in E\}$ incident on $v$, to obtain the graph $(V \backslash v, \mathcal{E} \backslash \mathcal{E}_v)$.

- **Edge deletion**: Given a graph $(V, \mathcal{E})$, an edge $E \in \mathcal{E}$ is *deleted* to obtain the graph $(V, \mathcal{E} \backslash E)$.

- **Edge contraction**: Given a graph $(V, \mathcal{E})$, an edge $\{u, v\} \in \mathcal{E}$ is *contracted* to form a single vertex $u'$ with edges to every vertex in $V \backslash \{u, v\}$ that previously had an edge to either $u$ or $v$. Thus, the resulting graph has one less vertex than the original graph.

Figure 2 gives an example of each of these operations. The graph $\mathcal{H}_1$ is a minor of $\mathcal{G}$, and is obtained from $\mathcal{G}$ by deleting the edge $\{5, 6\}$. Next, $\mathcal{H}_2$ is obtained from $\mathcal{H}_1$ by deleting the vertex 7, and the corresponding edges $\{4, 7\}, \{7, 8\}$ that are incident on 7. Thus, $\mathcal{H}_2$ is a minor of both $\mathcal{H}_1$ and $\mathcal{G}$. Finally, $\mathcal{H}_3$ is obtained from $\mathcal{H}_2$ by contracting the edge $\{5, 8\}$ to form the new vertex $8'$, which now has an edge to vertices 2, 4, and 9. The graph $\mathcal{H}_3$ is a minor of each of the graphs $\mathcal{G}$, $\mathcal{H}_1$, and $\mathcal{H}_2$.

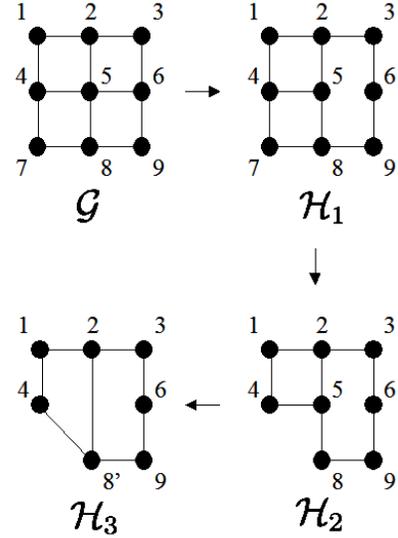

Figure 2: A graph $\mathcal{G}$, and three of its minors $\mathcal{H}_1$, $\mathcal{H}_2$, $\mathcal{H}_3$ obtained by edge deletion, followed by vertex deletion, and finally edge contraction.

In a series of over twenty papers, Robertson and Seymour investigated various aspects of graph minors and proved several important results in graph theory. The following theorem played a key role in proving many of these results; it provides a connection between treewidth and graph minors, and forms a critical component of our analysis.

**Theorem 2.1.** *(Robertson et al., 1994)* Let $\mathcal{G}$ be a $g \times g$ grid. There exists a finite $\kappa_{\mathrm{GM}}(g)$ such that $\mathcal{G}$ is a minor of all graphs with treewidth greater than $\kappa_{\mathrm{GM}}(g)$. Further, the best known bounds on $\kappa_{\mathrm{GM}}(g)$ are that $c_1 g^2 \log g \leq \kappa_{\mathrm{GM}}(g) \leq 2^{c_2 g^5}$, where $c_1$ and $c_2$ are universal constants (i.e. they are independent of $g$).

Thus, each grid-structured graph is a minor of all graphs with sufficiently large treewidth. Robertson et al. (1994) expressed the belief that $\kappa_{\mathrm{GM}}(g)$ is closer to $c_1 g^2 \log g$ than $2^{c_2 g^5}$, and may even be on the order of $g^2 \log g$. In addition, Demaine et al. (2006) build further support for this belief and conjecture that $\kappa_{\mathrm{GM}}(g) \sim g^3$. Consequently, we have the following grid-minor hypothesis.

**Grid-minor hypothesis**: $\kappa_{\text{GM}}(g)$, as defined in Theorem 2.1, is polynomial in $g$.

This hypothesis is a key assumption in the proof of our 'Main Result' as stated in the introduction. Next, we state a restricted result that relates graph minors and treewidth for planar graphs. A planar graph (Bollobás, 1998) is one that can be drawn on a plane with no two edges intersecting each other.

**Theorem 2.2.** *(Robertson et al., 1994) There exist universal constants $c_3$ and $c_4$ such that the following holds. Let $\mathcal{G}$ be a $g \times g$ grid. Then, (a) $\mathcal{G}$ is a minor of all planar graphs with treewidth greater than $c_3 g$. Further, (b) all planar graphs of size (number of vertices) less than $c_4 g$ are minors of $\mathcal{G}$.*

Hence, Theorem 2.2 states that $\kappa_{\text{GM}}(g)$ is actually *linear* in $g$ for planar graphs.

### 2.4.1 Inference and graph minors

Let $\mathcal{M}(\mathcal{G}, q)$ refer to the set of all possible choices for potential functions on the vertices and edges of $\mathcal{G} = (V, \mathcal{E})$, with the variables having maximum cardinality $q$. That is, each $\psi \in \mathcal{M}(\mathcal{G}, q)$ is specified as $\psi = \{\psi_v, v \in V\} \cup \{\psi_E, E \in \mathcal{E}\}$. In the following lemma, we relate the complexity of inference in a minor of a graph $\mathcal{G}$ to inference in $\mathcal{G}$.

**Lemma 2.3.** *Let $\mathcal{H}$ be a minor of $\mathcal{G}$, and let $\psi_{\mathcal{H}} \in \mathcal{M}(\mathcal{H}, q)$. There exists a $\psi_{\mathcal{G}} \in \mathcal{M}(\mathcal{G}, q)$ such that $Z(\psi_{\mathcal{H}}) = Z(\psi_{\mathcal{G}})$. Moreover, $\psi_{\mathcal{G}}$ can be computed in linear time given $\psi_{\mathcal{H}}$ and the sequence of minor operations that transform $\mathcal{G}$ to $\mathcal{H}$.*

*Proof.* All we need to show is that if a graph $\mathcal{H} = (V_{\mathcal{H}}, \mathcal{E}_{\mathcal{H}})$ is obtained from another graph $\mathcal{G} = (V_{\mathcal{G}}, \mathcal{E}_{\mathcal{G}})$ by just a *single* application of one of the standard minor operations, then we can transform a given $\psi_{\mathcal{H}} \in \mathcal{M}(\mathcal{H}, q)$ into a $\psi_{\mathcal{G}} \in \mathcal{M}(\mathcal{G}, q)$ with $Z(\psi_{\mathcal{G}}) = Z(\psi_{\mathcal{H}})$.

**Vertex deletion**: Suppose that $v \in V_{\mathcal{G}}$ as well as edges $\mathcal{E}_v \subseteq \mathcal{E}_{\mathcal{G}}$ that are incident on $v$ in $\mathcal{G}$ are deleted. Let $\psi_v = \frac{1}{q}$ and let $\psi_E = 1, \forall E \in \mathcal{E}_v$. Letting $\psi_{\mathcal{G}} = \cup_{E \in \mathcal{E}_v} \psi_E \cup \psi_v \cup \psi_{\mathcal{H}}$, one can check that $Z(\psi_{\mathcal{G}}) = Z(\psi_{\mathcal{H}})$.

**Edge deletion**: Suppose that $E \in \mathcal{E}_{\mathcal{G}}$ is deleted. Setting $\psi_E = 1$, and $\psi_{\mathcal{G}} = \psi_{\mathcal{H}} \cup \psi_E$, one can check that $Z(\psi_{\mathcal{G}}) = Z(\psi_{\mathcal{H}})$.

**Edge contraction**: Suppose that $\{u, v\} \in \mathcal{E}_{\mathcal{G}}$ is contracted to form the new vertex $u' \in V_{\mathcal{H}}$. We define $\psi_{\{u,v\}}(x_u, x_v) = \delta(x_u - x_v)$, where $\delta(\cdot)$ is the Kronecker delta function that evaluates to 1 if the argument is 0, and 0 otherwise. For the edge potentials, if a vertex $w \in V_{\mathcal{G}} \setminus \{u, v\}$ is originally connected in $\mathcal{G}$ by an edge to only one of $u$ or $v$, then we set the corresponding $\psi_{\{u,w\}}$ or $\psi_{\{v,w\}}$ to be equal to $\psi_{\{u',w\}}$. If both $u$ and $v$ are originally connected by edges to $w$ in $\mathcal{G}$, then we define $\psi_{\{u,w\}} = \psi_{\{u',w\}}$ and $\psi_{\{v,w\}} = 1$. Finally, we define the vertex potentials as $\psi_u = \psi_{u'}$ and $\psi_v = 1$. Letting all the other vertex and edge potentials in $\mathcal{G}$ be the same as those in $\mathcal{H}$, it is easily seen that $Z(\psi_{\mathcal{G}}) = Z(\psi_{\mathcal{H}})$. □

Thus, an inference problem in a minor of $\mathcal{G}$ can be transformed to an inference problem in $\mathcal{G}$. Consequently, this result allows us to establish hardness of inference in a graph $\mathcal{G}$, by establishing hardness of inference in a minor of $\mathcal{G}$.

## 3 PROBLEM STATEMENT

Let $T_f(I)$ denote the runtime of an algorithm $f$ on input $I$. We consider inference algorithms that take as input a graph $\mathcal{G} = (V, \mathcal{E})$ and an element of $\mathcal{M}(\mathcal{G}, q)$ (i.e., potentials defined with respect to the vertices and edges of $\mathcal{G}$), and compute the partition function $Z(\psi)$. We would like to investigate the impact of the treewidth $\text{tw}(\mathcal{G})$ of the graph $\mathcal{G}$ on the required runtime of any inference algorithm.

Typical complexity analysis studies the worst case, or maximum, runtime of an algorithm over all inputs. Since inference in a graphical model is **NP**-hard, and assuming $\mathbf{NP} \neq \mathbf{P}$, we know that the worst case runtime of any inference algorithm must scale super-polynomially with the size of the graph. That is, the *maximum* runtime over all graphs is super-polynomial.

Our focus in this paper is on studying the following "best case" complexity of inference:

$$\beta_f(k, q) = \min_{\mathcal{G}: \text{tw}(\mathcal{G}) = k} \max_{\psi \in \mathcal{M}(\mathcal{G}, q)} T_f(\mathcal{G}, \psi). \quad (3)$$

In words, $\beta_f(k, q)$ captures the complexity of inference as a function of treewidth by finding the "best", or "easiest" graph of treewidth $k$ for each $k$. Since we are primarily concerned with bounds that are independent of the cardinality $q$, we will specifically consider the case $q = 2$ and define $\mathcal{M}(\mathcal{G}) = \mathcal{M}(\mathcal{G}, 2), \beta_f(k) = \beta_f(k, 2)$.

**Main Question**: How does $\beta_f(k)$, as defined in (3), grow as a function of the treewidth $k$ for any inference algorithm $f$? Does there exist an inference algorithm $f$ for which $\beta_f(k)$ grows only polynomially with $k$?

If there exists an $f$ such that $\beta_f(k)$ is polynomial in $k$, then there exists a class of structures with unbounded treewidth in which inference would be tractable. Alternatively, if $\beta_f(k)$ is not polynomial in $k$ for any procedure $f$, then bounding the treewidth is the only structural restriction on graphical models that leads to tractable inference.

The quantity $\beta_f(k)$ in the 'Main Question' refers to a uniform algorithm, i.e. a single algorithm that should work for graphs of all treewidths. However, to answer this question we will actually study a slightly harder question, where we allow non-uniform algorithms specialized to a sequence of graphs of increasing treewidths. Given a sequence of graphs $\{\mathcal{G}_k\}_{k=1}^\infty$ with $\mathrm{tw}(\mathcal{G}_k) = k$, we will analyze the runtime of any (non-uniform) sequence $f = \{f_k\}_{k=1}^\infty$ of algorithms (i.e. a "non-uniform algorithm"), where $f_k$ solves the inference problem on $\mathcal{G}_k$. For any such sequence, we study how the runtime increases (taking worst case over potential functions) with $k$, i.e. $\max_{\psi \in \mathcal{M}(\mathcal{G}_k)} T_{f_k}(\mathcal{G}_k, \psi)$ as a function of $k$. Taking the infimum over the choice of the sequence of graphs $\{\mathcal{G}_k\}_{k=1}^\infty$ (i.e. choosing the "easiest" sequence of graphs of increasing treewidth) gives a lower bound on $\beta_f(k)$.

Our 'Main Question' pertains to *exact* inference. We also investigate the tractability of obtaining an approximation to the partition function. Specifically, we consider the problem of computing $Z(\psi)$ up to an additive constant $\varepsilon$, perhaps using a randomized procedure. Focusing on (randomized) algorithms $f(\mathcal{G}, \psi, \varepsilon)$ that provide a $\widehat{Z}$ such that $Z(\psi) - \varepsilon \leq \widehat{Z} \leq Z(\psi) + \varepsilon$ (with high probability—see Section 5.2), we consider:

$$\beta_f^\varepsilon(k) = \min_{\mathcal{G}:\mathrm{tw}(\mathcal{G})=k} \max_{\psi \in \mathcal{M}(\mathcal{G})} T_f(\mathcal{G}, \psi, \varepsilon). \qquad (4)$$

Note that $\mathcal{M}(\mathcal{G}) = \mathcal{M}(\mathcal{G}, 2)$. We can now ask a question analogous to our 'Main Question', for $\beta_f^\varepsilon(k)$ rather than $\beta_f(k)$.

## 4 CONSTRAINT SATISFACTION AND INFERENCE

A *constraint satisfaction problem* (CSP) is defined as a set of constraints specified on subsets of a collection of discrete-valued variables. Each constraint is said to be *satisfied* for some stipulated configurations of the variables in the constraint. The problem is to identify a configuration of the variables that satisfies all the constraints (i.e., find a *satisfying* assignment). We will consider CSP as a decision problem — the problem of deciding if such a satisfying assignment exists. We will mostly be concerned with 2-CSPs: CSPs in which each constraint involves only two variables. Note that one can associate a graph with an instance of a 2-CSP, with the vertices representing the variables and edges present only between those vertices that appear in the same constraint. A related problem is the MAX CSP in which one is interested in configurations of the variables that *maximize* the number of satisfied constraints. Again, we will refer to MAX CSP as the problem of deciding, for some integer $d$, if there are any configurations that simultaneously satisfy more that $d$ constraints.

An important special case of a CSP is the SAT problem, in which disjunctive constraints are specified on binary variables. Although polynomial time algorithms exist for 2-SAT, the MAX 2-SAT problem is **NP**-complete. In fact we have that planar MAX 2-SAT, in which instances are restricted to those defined on planar graphs, is also **NP**-complete (Guibas et al., 1991).

To obtain sharper results, we will use the so-called "Exponential-Time Hypothesis" (Impagliazzo et al., 2001):

**Exponential-time hypothesis (ETH)**: There exists no non-uniform algorithm[2] that can solve arbitrary instances of $n$-variable 3-SAT in time $2^{o(n)}$.

Note that **NP** $\not\subseteq$ **P/poly** would merely state that there exists no polynomial-time (non-uniform) algorithm for arbitrary $n$-variable instances of 3-SAT (since 3-SAT is **NP**-complete). Thus, the ETH is a stronger assumption than **NP** $\not\subseteq$ **P/poly**, and consequently, allows one to obtain sharper bounds on the growth of $\beta_f(k)$ and $\beta_f^\varepsilon(k)$ (see Section 5 for more details).

In order to translate hardness results for CSPs and MAX CSPs to the problem of inference in graphical models, we prove the following lemma that transforms instances of 2-CSPs to inference problems in graphical models. Specifically, we show that each instance of a MAX 2-CSP can be mapped to a particular decision-version of an inference problem.

**Lemma 4.1.** *Let $I = (x_1, \cdots, x_n; \mathcal{R})$ be an instance of a MAX 2-CSP problem, where $x_1, \cdots, x_n$ are discrete-valued variables (of cardinality $q$) and $\mathcal{R}$ is a set of constraints. Let $\mathcal{G} = (V, \mathcal{E})$ denote the graph which represents the instance $I$. There exist a set of potentials $\psi \in \mathcal{M}(\mathcal{G}, q)$ and a function $h : \{0, \cdots, |\mathcal{R}|\} \to \mathbb{R}^+$ with the following property[3]: at least $d$ disjunctions in $\mathcal{R}$ can be satisfied simultaneously if and only if $Z(\psi) \geq h(d)$. Moreover, the construction of the potentials $\psi$ and the evaluation of the function $h$ are polynomial-time operations, given $I$.*

*Proof.* Let $\mathcal{G} = (V, \mathcal{E})$ denote the graph which represents the instance $I$. Hence, $|V| = n$ with each variable being assigned to a vertex and $\mathcal{E}$ contains only those pairs of vertices for which the corresponding variables

---

[2] Impagliazzo et al. (2001) refer to the uniform version of ETH, but their results equally apply to the above-stated non-uniform version of the hypothesis, which is also widely believed to be true.

[3] $|\mathcal{R}|$ denotes the number of constraints in $\mathcal{R}$.

appear in the same relation, so that $|\mathcal{E}| = |\mathcal{R}|$. For each $E \in \mathcal{R}$, define

$$\psi_E(x_E) = \begin{cases} 1, & x_E \text{ satisfies } E \\ \varepsilon, & \text{otherwise.} \end{cases}$$

Define vertex potentials similarly for each vertex constraint, and set $\psi_v = 1$ for other vertices. Choose $\varepsilon \in (0, \frac{1}{q^n})$. Let $h(d) = \varepsilon^{|\mathcal{E}|-d}$. If $I$ is such that at least $d$ constraints can be simultaneously satisfied, it is clear that $Z(\psi) \geq \varepsilon^{|\mathcal{E}|-d}$. Alternatively, if $I$ is such that $d$ or more constraints can never be satisfied simultaneously, then we have that

$$Z(\psi) \leq \sum_{x_V \in \{0, \cdots, q-1\}^n} \varepsilon^{|\mathcal{E}|-d+1} = q^n \varepsilon h(d) < h(d).$$

$\square$

By setting $d = |\mathcal{R}|$ in the above lemma, one can transform instances of a 2-CSP problem to a decision-version of an inference problem. Next, we translate the recent result in (Marx, 2007) for 2-CSPs to a complexity result for inference.

**Theorem 4.2.** *(Marx, 2007)*[4] *Let $\{\mathcal{G}_k\}_{k=1}^{\infty}$ be any sequence of graphs indexed by treewidth. Suppose that there exists an algorithm $g$ for instances of 2-CSPs, with variables of arbitrary cardinality, defined on the graphs $\mathcal{G}_k$. Let $q(\psi)$ be the maximum cardinality of a variable referred to by the constraints $\psi$. If $T_g(\mathcal{G}_k, \psi) = q(\psi)^{o(\frac{k}{\log k})}$, then the ETH fails.*

**Corollary 4.3.** *Let $f$ be any algorithm that can perform inference on graphical models with variables of arbitrary cardinality. Under the ETH, for any $r(k) = o(k/\log k)$ there exist $q, k$ such that $\beta_f(k, q) > q^{r(k)}$.*

A consequence of this corollary is that the junction-tree algorithm (Cowell et al., 1999), which scales as $q^k$, is in a sense near-optimal (assuming the ETH). However, as we noted in the introduction, this result has a significant weakness in that it provides an asymptotic lower bound only for sufficiently large cardinalities. It does not provide a lower bound for any fixed cardinality $q$. This restriction plays an important role in the reductions in (Marx, 2007), in which large sets of variables in an intermediate model are represented using a single high-cardinality variable. In the following section, we describe our main results for the complexity of inference in graphical models with *binary* variables, which are typically of most interest to the machine learning community.

Another important class of problems arising from constraint satisfaction is that of *counting* the number of satisfying assignments in a CSP. Such counting problems are titled #CSPs, and planar #2-SAT falls under the class of #**P**-complete problems (Vadhan, 2001). Instances of these problems can also be transformed to inference in graphical models.

**Lemma 4.4.** *Let $I = (x_1, \ldots, x_n; \mathcal{R})$ be an instance of a #2-CSP problem, where $x_1, \ldots, x_n$ are discrete-valued variables of cardinality $q$ and $\mathcal{R}$ is a set of constraints. Let $\mathcal{G}$ be the graph which represents the instance $I$. There exists a set of potentials $\psi \in \mathcal{M}(\mathcal{G}, q)$ such that the number of satisfying assignments is $\lfloor Z(\psi) \rfloor$. Moreover, the construction of the potentials $\psi$ is a polynomial-time operation, given $I$.*

*Proof.* The construction of the potentials is similar to that in the proof of Lemma 4.1. One can check that the resulting partition function has the property that the integer part is equal to the number of satisfying solutions. $\square$

In the following section, we use the #**P**-completeness of planar #2-SAT to demonstrate the hardness of approximation of $Z(\psi)$ up to an additive constant.

## 5 MAIN RESULTS

We present our main results for graphical models with binary-valued variables in this section.

### 5.1 EXACT INFERENCE

**Theorem 5.1.** *Let $\{\mathcal{G}_k\}_{k=1}^{\infty}$ be an infinite sequence of graphs indexed by treewidth. Let $f = \{f_k\}_{k=1}^{\infty}$ be any (possibly non-uniform) sequence of algorithms that solves the inference problem on $\{\mathcal{G}_k\}$ with binary variables. Furthermore, let $T(k)$ denote the worst-case running time of $f$ on $\mathcal{G}_k$ (i.e, $T(k) = \max_{\psi \in \mathcal{M}(\mathcal{G}_k)} T_{f_k}(\mathcal{G}_k, \psi)$).*

*(a) Assuming that $\mathbf{NP} \nsubseteq \mathbf{P/poly}$ and that the grid-minor hypothesis holds, $T(k)$ is super-polynomial in $k$. Hence, $\beta_f(k)$ as defined in (3) is super-polynomial with respect to $k$.*

*(b) Assuming that $\kappa_{\text{GM}}(g) = O(g^r)$ in the grid-minor hypothesis and the ETH, we have that $T(k) = 2^{\Omega(k^{1/2r})}$. Hence, $\beta_f(k) = 2^{\Omega(k^{1/2r})}$.*

*Proof.* (a) Suppose (for contradiction) that there exists a (possibly non-uniform) polynomial time algorithm $f$ that solves the inference problem on $\{\mathcal{G}_k\}_{k=1}^{\infty}$. More precisely, let $f = \{f_k\}_{k=1}^{\infty}$ be a sequence of algorithms such that $f_k$ solves the inference problem on $\mathcal{G}_k$ in polynomial time. Assuming the grid-minor hypothesis, we will demonstrate that this implies a non-uniform polynomial time algorithm for the inference problem on any planar graph. Recall that planar MAX 2-SAT is **NP**-complete (Guibas et al., 1991) and polynomial-time reducible to the inference problem on planar graphs (Lemma 4.1). This provides a (non-uniform) polynomial time algorithm for an **NP**-complete problem, contradicting the $\mathbf{NP} \not\subseteq \mathbf{P/poly}$ assumption.

---

[4] The statement here is actually of a non-uniform variant of the result of Marx (2007).

Given an instance $(\mathcal{G}, \psi)$ of the inference problem on planar graphs, we proceed as follows: Let $|\mathcal{G}| = s$. By Theorem 2.2, $\mathcal{G}$ is a minor of the $s/c_4 \times s/c_4$ grid. Furthermore, the sequence of minor operations that transform a $s/c_4 \times s/c_4$ grid to $\mathcal{G}$ can be obtained in polynomial time (Tamassia & Tollis, 1989). Thus, using Lemma 2.3, the inference problem $(\mathcal{G}, \psi)$ can be reduced to an inference problem on the $s/c_4 \times s/c_4$ grid in time linear in $s$. By Theorem 2.1, the $s/c_4 \times s/c_4$ grid is a minor of $\mathcal{G}_{\kappa_{\text{GM}}(s/c_4)}$. We will now use as "non-uniform advice" the sequence of minor operations that transform $\mathcal{G}_{\kappa_{\text{GM}}(s/c_4)}$ to the $s/c_4 \times s/c_4$ grid. Note that this depends only on the input size $s$ and not on the actual instance $(\mathcal{G}, \psi)$. Using Lemma 2.3 again, we can reduce the inference problem on the $s/c_4 \times s/c_4$ grid to an inference problem on $\mathcal{G}_{\kappa_{\text{GM}}(s/c_4)}$ in linear time. We now use $f_{\kappa_{\text{GM}}(s/c_4)}$ to solve the inference problem on $\mathcal{G}_{\kappa_{\text{GM}}(s/c_4)}$, thus solving the original inference problem $(\mathcal{G}, \psi)$. The fact that $T(k)$, and thus also the size of the graph $\mathcal{G}_k$, is at most polynomial in $k$ and the grid-minor hypothesis (i.e, $\kappa_{\text{GM}}(g) = \text{poly}(g)$) imply that the above algorithm is a polynomial time (non-uniform) algorithm for the inference problem on planar graphs.

(b) We obtain the tighter hardness result by carefully analyzing the running time of the inference algorithm on planar graphs suggested in (a). It can be easily checked that the above algorithm runs in time $T(\kappa_{\text{GM}}(s/c_4))$, which is $T(O(s^r))$ if $\kappa_{\text{GM}}(g) = O(g^r)$. Combining this with the reduction from 3-SAT to planar MAX 2-SAT ((Litchenstein, 1982; Guibas et al., 1991)), which blows up the instance size by a quadratic factor, we obtain a $T(O(n^{2r}))$ time non-uniform algorithm for $n$-variable instances of 3-SAT. Recall that the ETH states there exists no (non-uniform) algorithm for arbitrary $n$-variable instances of 3-SAT that has running time $2^{o(n)}$. Hence, assuming the grid-minor hypothesis and the ETH, we must have that $T(O(n^{2r}))$ is at least $2^{O(n)}$ or equivalently that $T(n)$ is at least $2^{\Omega(n^{1/2r})}$. □

Theorem 5.1 provides an answer to the 'Main Question' in Section 3, and comprises the 'Main Result' described in the introduction. Notice that the ETH assumption enables a sharper performance bound instead of the simpler result that $\beta_f(k)$ is super-polynomial in $k$ (of part (a)). Next, we have the following theorem for planar graphs that does *not* require assumption of the grid-minor hypothesis. Define

$$\beta_f^{\text{planar}}(k) = \min_{\mathcal{G}:\text{tw}(\mathcal{G})=k,\ \mathcal{G}\text{ planar}} \max_{\psi \in \mathcal{M}(\mathcal{G})} T_f(\mathcal{G}, \psi). \quad (5)$$

**Theorem 5.2.** *Let $f$ be any inference algorithm that operates on graphical models with binary variables defined on planar graphs.*

*(a) Assuming that $\mathbf{NP} \not\subseteq \mathbf{P/poly}$, $\beta_f^{\text{planar}}(k)$ as defined in (5) is super-polynomial with respect to $k$.*

*(b) Under the ETH, $\beta_f^{\text{planar}}(k) = 2^{\Omega(k^{1/2})}$.*

*Proof.* The proof is similar to that of Theorem 5.1, but the grid-minor hypothesis is not required due to Theorem 2.2. □

Based on the results in (Demaine et al., 2005) Theorem 5.2 holds more generally for the inference problem in graphical models defined on bounded-genus graphs, of which planar graphs are a special case.

### 5.2 APPROXIMATE INFERENCE

The above results show that exact inference is intractable in any class of graphs with unbounded treewidth. Here, we prove that even obtaining an approximation within some additive constant to the partition function is intractable. Our result uses Lemma 4.4 along with the fact that planar #2-SAT is #**P**-complete (Vadhan, 2001).

**Theorem 5.3.** *Suppose that the grid-minor hypothesis holds. Let $f$ be any (randomized) approximate inference algorithm that operates on graphical models with binary variables. Let $\varepsilon > 0$ and $0 < \delta < 1$ be specified, and suppose that $f(\mathcal{G}, \psi, \varepsilon)$ provides an approximation $\widehat{Z}$ such that*

$$\Pr[Z(\psi) - \varepsilon \leq \widehat{Z} \leq Z(\psi) + \varepsilon] \geq 1 - \delta.$$

*If $\beta_f^\varepsilon(k)$, as defined in (4), is polynomial in $k$, $\frac{1}{\varepsilon}$, and $\log(\frac{1}{\delta})$, then $\mathbf{NP} \subseteq \mathbf{P/poly}$.*

*Proof.* Unless $\mathbf{NP} \subseteq \mathbf{P/poly}$, there is no randomized non-uniform procedure that can approximate the solution to a #**P**-complete problem to within a constant $c$ with probability greater than $1 - \delta$, which is polynomial in the size of the problem, $c$, and $\log(\frac{1}{\delta})$ (Vazirani, 2004). Based on Lemma 4.4, we have that instances of planar #2-SAT can be reduced to performing inference in a model defined on a planar graph so that the number of solutions is equal to $\lfloor Z(\psi) \rfloor$. Using the fact that planar #2-SAT is #**P**-complete, one can prove this result by following the same line of analysis adopted in the proof of Theorem 5.1. □

# 6 CONCLUSION

With increasing interest in understanding various inference procedures and providing conditions under which they are correct and tractable, it is important to understand whether there might indeed be some structural property, other than treewidth, which can guarantee tractable inference. In this paper we studied this issue, presenting and discussing the relevant literature from the CSP community as well as relevant graph theory concepts, and can conclude that it is not likely such an alternate property exists—finding a property that ensures tractability of inference without bounding treewidth would imply providing a counterexample to Robertson and Seymour's grid-minor hypothesis.

We believe that relating the "best case" complexity of inference to the grid-minor hypothesis provides substantial evidence that inference remains hard even in the "easiest" high-treewidth graph structures. Nevertheless, it would be of great interest to prove the results in this paper without resorting to the grid minor hypothesis. It would also be useful to obtain hardness results that are valid even for reasonably restricted classes of potential functions, e.g. potential functions with bounded dynamic range.

**Acknowledgements** We would like to thank Lance Fortnow and Jaikumar Radhakrishnan for helpful discussions and referring us to (Tamassia & Tollis, 1989).